\documentclass[format=acmsmall, review=false, screen=true]{acmart}

\usepackage{soul}
\usepackage{tcolorbox}
\usepackage{enumitem}
\usepackage{tikz}
\usetikzlibrary{mindmap,backgrounds}

\newcommand{\find}[1]{
\begin{tcolorbox}[leftrule=0.5mm,toprule=0mm,bottomrule=0mm,left=1pt,right=1pt,top=2pt,bottom=2pt]
\em #1
\end{tcolorbox}
}

\AtBeginDocument{%
  }

\acmJournal{TOSEM}

\begin{document}

\title{Efficient and Green Large Language Models for Software Engineering: Literature Review, Vision, and the Road Ahead}

\author{Jieke Shi}
\orcid{0000-0002-0799-5018}
\email{jiekeshi@smu.edu.sg}
\author{Zhou Yang}
\orcid{0000-0001-5938-1918}
\email{zyang@smu.edu.sg}
\authornote{Corresponding author}
\author{David Lo}
\orcid{0000-0002-4367-7201}
\email{davidlo@smu.edu.sg}

\affiliation{%
  \institution{Singapore Management University}
  \country{Singapore}
}

\begin{abstract}
  Large Language Models (LLMs) have recently shown remarkable capabilities in various software engineering tasks, spurring the rapid growth of the Large Language Models for Software Engineering (LLM4SE) area. However, limited attention has been paid to developing efficient LLM4SE techniques that demand minimal computational cost, time, and memory resources, as well as green LLM4SE solutions that reduce energy consumption, water usage, and carbon emissions.

This paper aims to redirect the focus of the research community towards the efficiency and greenness of LLM4SE, while also sharing potential research directions to achieve this goal. It commences with a brief overview of the significance of LLM4SE and highlights the need for efficient and green LLM4SE solutions. Subsequently, the paper presents a vision for a future where efficient and green LLM4SE revolutionizes the LLM-based software engineering tool landscape, benefiting various stakeholders, including industry, individual practitioners, and society. The paper then delineates a roadmap for future research, outlining specific research paths and potential solutions for the research community to pursue. While not intended to be a definitive guide, the paper aims to inspire further progress, with the ultimate goal of establishing efficient and green LLM4SE as a central element in the future of software engineering.
\end{abstract}

\begin{CCSXML}
  <ccs2012>
     <concept>
         <concept_id>10002944.10011122.10002945</concept_id>
         <concept_desc>General and reference~Surveys and overviews</concept_desc>
         <concept_significance>500</concept_significance>
         </concept>
     <concept>
         <concept_id>10011007.10011074.10011092</concept_id>
         <concept_desc>Software and its engineering~Software development techniques</concept_desc>
         <concept_significance>500</concept_significance>
         </concept>
     <concept>
         <concept_id>10010147.10010178</concept_id>
         <concept_desc>Computing methodologies~Artificial intelligence</concept_desc>
         <concept_significance>500</concept_significance>
         </concept>
   </ccs2012>
\end{CCSXML}

\ccsdesc[500]{General and reference~Surveys and overviews}
\ccsdesc[500]{Software and its engineering~Software development techniques}
\ccsdesc[500]{Computing methodologies~Artificial intelligence}

\keywords{Software Engineering, Large Language Models, Efficiency, Greenness}

\maketitle

\section{Introduction}
\label{sec:intro}

Recent advances in Large Language Models (LLMs) have spurred their widespread adoption in various fields, notably within software engineering~\cite{hou2023large,fan2023large, zhang2023survey,zheng2023survey, wang2024software,gormez2024large,belzner2023large}. Trained on massive amounts of data, LLMs have shown their prowess in processing and generating code written in various programming languages, with exceptional performance across a spectrum of code-related tasks, ranging from code generation~\cite{xu2022systematic,chen2021evaluating,codesuggestion} and summarization~\cite{ahmed2022few,ahmed2024automatic,li2024machines} to vulnerability detection~\cite{zhou2024large,purba2023software,thapa2022transformer}, program repair~\cite{jiang2023impact,jin2023inferfix,xia2023automated}, and more~\cite{schafer2023empirical,zhang2023cupid,niu2024fair,feng2024prompting, shi2022can}. This success has fueled the rapid growth of the Large Language Models for Software Engineering (LLM4SE) area, which has recently become one of the most active areas in software engineering.

To date, certain LLM4SE techniques have been translated into real-world applications. For instance, GitHub Copilot~\cite{githubGitHubCopilot}, powered by OpenAI's GPT-4~\cite{achiam2023gpt}, assists developers in writing code, while Google VirusTotal Code Insight~\cite{virustotalIntroducingVirusTotal} leverages the Sec-PaLM model~\cite{googleGoogleCloud} for malware analysis. These tools have introduced a new level of automation to software development, enhancing both developer productivity~\cite{githubResearchQuantifying,coutinho2024role} and software quality~\cite{10.1145/3643762,githubResearchQuantifyingquality}. However, due to the computationally-intensive and energy-demanding nature of LLMs~\cite{hoffmann2022empirical,li2024toward,samsi2023words}, most of these LLM4SE applications are facilitated by large companies with abundant computational resources and energy supplies that are not readily available to the broader software engineering community, including startups and individual developers. To democratize LLM4SE techniques and make them beneficial to all software engineering practitioners, several key issues must be addressed, particularly their unsatisfactory efficiency (i.e., high computational cost, memory usage, and time consumption) and lack of greenness (i.e., high energy usage, water consumption, and carbon emissions)~\cite{yang2024harnessing,li2023making}.

Our research community has recently acknowledged the growing need to enhance LLM4SE techniques' efficiency and greenness, with several studies making strides for this purpose, such as those by Shi et al.~\cite{compressor,avatar} and Sun et al.~\cite{10.1145/3688831,10.1145/3650212.3680347}. However, this area remains far from fully explored, with many breakthroughs yet to be made, and the overall landscape still unclear as we look toward 2030 and beyond. This paper, therefore, presents our vision for the future of efficient and green LLM4SE, identifies research gaps and potential opportunities, and outlines a roadmap for our research community to advance this area. Our vision is encapsulated in the following statement:

% \begin{minipage}{0.8\textwidth}
%     \begin{center}
%         {\it "Efficient and green LLM4SE will revolutionize the LLM-based software engineering tool landscape, benefiting various stakeholders, including industry, individual practitioners, and society."}
%     \end{center}
% \end{minipage}

\begin{tcolorbox}[width=\textwidth, box align=center, left=0.5cm, top=0pt, bottom=0pt, right=0.5cm, colback=white, frame empty]
    % \begin{center}
        {\it ``By 2030, LLM4SE tools will be efficient, green, more affordable, and more accessible to individual software practitioners, while fostering sustainability and reducing the environmental impact across the software industry and society.''}
    % \end{center}
\end{tcolorbox}

% This vision is elaborated in detail in~\autoref{sec:vision}. In~\autoref{sec:current_contributions}, we review key historical milestones and recent advances in efficient and green LLM4SE techniques, while~\autoref{sec:road} provides a roadmap for future research, both to validate our vision. This paper aims to answer the following research questions:
% {\bf RQ1}: How do current efficient and green LLM4SE techniques contribute to the area, and are they sufficient?
% {\bf RQ2}: What might the future landscape of LLM4SE tools look like in 2030 and beyond, with improved efficiency and greenness?
% {\bf RQ3}: How can our research community advance towards more efficient and green LLM4SE in the coming decade?

The rest of the paper is organized as follows. In~\autoref{sec:background}, we define efficient and green LLM4SE and highlight their importance.~\autoref{sec:current_contributions} then reviews on the current state of LLM4SE solutions in terms of efficiency and greenness, followed by a vision for the future of efficient and green LLM4SE in~\autoref{sec:vision}. We provide a roadmap outlining specific research paths and potential solutions that the research community can pursue in~\autoref{sec:road}. Finally, we conclude with a summary of the key points and a call to action for the research community to embrace efficient and green LLM4SE techniques in~\autoref{sec:conclusion}.

% The other sections of this paper are organized as follows.~\autoref{sec:background} presents the definition of efficient and green LLM4SE and highlights their significance. while~\autoref{sec:conclusion} concludes with a summary of the key points and a call to action for the research community to embrace efficient and green LLM4SE techniques.

% The rest of the paper is organized as follows. We reflect on the current state of LLM4SE solutions in terms of efficiency and greenness in Section~\ref{sec:current_contributions}, and then present a vision for the future of efficient and green LLM4SE in Section~\ref{sec:vision}. Section~\ref{sec:road} proposes a roadmap to address the above critical needs, aiming to inspire further research and innovation. Section~\ref{sec:conclusion} concludes the paper with a summary of the key points and a call to action for the research community to embrace efficient and green LLM4SE solutions.
\section{Preliminaries}
\label{sec:background}

\subsection{Large Language Models (LLMs)}

Large Language Models (LLMs) refer to a class of neural network models that leverage the Transformer architecture~\cite{vaswani2017attention} to process and generate language text. The term ``large'' signifies the vast number of parameters in these models, typically ranging from hundreds of millions to billions. Here, we briefly introduce the mainstream LLMs that have driven the rapid progress in the LLM4SE area. Following Hou et al.~\cite{hou2023large}, we categorize these models into three groups based on their architectures: 1) encoder-only, 2) encoder-decoder, and 3) decoder-only LLMs. We also refer readers to recent surveys~\cite{hou2023large,yang2024robustness} for a more comprehensive overview of LLMs.

\subsubsection{Encoder-Only LLMs.} Encoder-only LLMs, such as BERT~\cite{devlin-etal-2019-bert} and RoBERTa~\cite{RoBERTa}, utilize only the encoder component of the Transformer architecture~\cite{vaswani2017attention} to process input and capture contextual information for downstream tasks. In the software engineering domain, CodeBERT~\cite{feng2020codebert} and GraphCodeBERT~\cite{GraphCodeBERT} are prominent examples of encoder-only LLMs tailored for programming languages. Additional models developed for various SE tasks include CuBERT~\cite{CuBERT}, CCBERT~\cite{zhou2023ccbert}, BERTOverflow~\cite{tabassum-etal-2020-code}, SOBERT~\cite{he2024representation}, CodeRetriever~\cite{li2022coderetriever}, FAIR~\cite{niu2024fair}, UniXcoder~\cite{guo2022unixcoder}, and VarBERT~\cite{10646727}.

\subsubsection{Encoder-Decoder LLMs.} Encoder-decoder LLMs are equipped with both the encoder and decoder components of Transformer~\cite{vaswani2017attention}, which can respectively process input and generate output text. These LLMs are more versatile than encoder-only models in that they can better handle a wide range of generation tasks. Models such as PLBART~\cite{plbart}, CodeT5~\cite{wang2021codet5}, CodeT5+~\cite{wang2023codet5+}, AlphaCode~\cite{li2022competition}, SPT-Code~\cite{10.1145/3510003.3510096}, CoTexT~\cite{phan2021cotext}, NatGen~\cite{chakraborty2022natgen}, CoditT5~\cite{zhang2022coditt5}, and GrammarT5~\cite{10.1145/3597503.3639125} showcase the effectiveness of encoder-decoder LLMs in various SE tasks.

\subsubsection{Decoder-Only LLMs.} Decoder-only LLMs, currently the most prevalent type of LLMs, are exemplified by OpenAI's GPT series~\cite{achiam2023gpt,brown2020languagemodelsfewshotlearners}. These models rely solely on the decoder component of the Transformer and are designed for autoregressive generation tasks, where they predict the next token in a sequence based on the preceding tokens. Numerous decoder-only LLMs tailored for code generation have been developed, including CodeGPT~\cite{CodeXGLUE}, GPT-C~\cite{svyatkovskiy2020intellicode}, Codex~\cite{chen2021evaluating}, PolyCoder~\cite{xu2022systematic}, CodeGen~\cite{codegen,nijkamp2023codegen2}, Code Llama~\cite{roziere2023code}, StarCoder~\cite{li2023starcoder,lozhkov2024starcoder}, and Magicoder~\cite{wei2024magicoder}.

\subsection{Definition of Efficient and Green LLM4SE}

\begin{figure}
    \centering
    \includegraphics[width=0.7\textwidth]{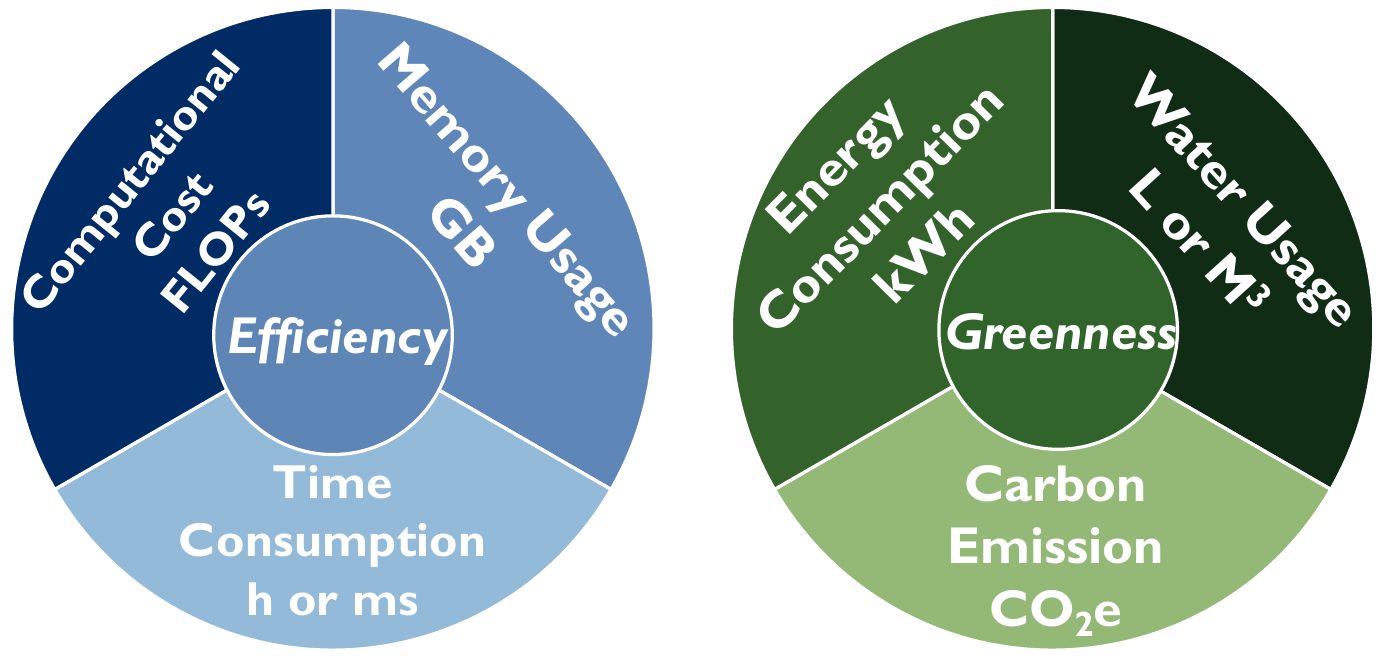}
    \caption{Six Dimensions of Efficient and Green LLM4SE, along with their metrics.}
    \label{fig:efficient_green_llm4se}
\end{figure}

As illustrated in \autoref{fig:efficient_green_llm4se}, this paper defines efficient and green LLM4SE techniques as those that minimize six key dimensions from training to deployment: computational cost, memory usage, and time consumption for efficiency, and energy consumption, water usage, and carbon emissions for greenness. These dimensions and their metrics are described as follows.

\vspace{0.08cm}
\noindent\textbf{Dimensions of Efficiency.} Computational cost refers to the total arithmetic operations (e.g., addition, multiplication) required for training or inference in LLM4SE techniques, typically measured in floating-point operations (FLOPs), as seen in recent studies~\cite{avatar,compressor}. Memory usage measures the amount of memory consumed during training or deployment, commonly expressed in gigabytes (GB). Time consumption accounts for the duration needed for training or the latency for inference, typically quantified in hours (h) for training and milliseconds (ms) for inference~\cite{10.1145/3611643.3616302,avatar,compressor}.

\vspace{0.08cm}
\noindent\textbf{Dimensions of Greenness.}
The term ``green'' is often used interchangeably with ``sustainable'' in the literature. However, recent studies~\cite{jarvenpaa2024synthesis,guldner2024development}, including work by Calero et al.~\cite{calero2024addressing} and Cruz et al.~\cite{cruz2024innovating}, have called for a distinction between the two, with ``green'' specifically referring to environmental sustainability, while ``sustainable'' covers a broader spectrum, including social and economic dimensions. In this paper, we focus on the environmental aspect of LLM4SE techniques and use the term ``green''; other aspects of sustainability are beyond our scope.

We define green LLM4SE techniques as those with low energy consumption, water usage, and carbon emissions. Energy consumption refers to the electrical power required during both training and inference phases, typically measured in kilowatt-hours (kWh)~\cite{avatar}. Water usage denotes the volume of water used for electricity generation or server cooling, often measured in liters (L) or cubic meters (m\textsuperscript{3})~\cite{li2023making}. Carbon emissions account for the release of greenhouse gases, primarily carbon dioxide (CO\textsubscript{2}), during training or inference, typically measured in kilograms (kg) or tons (t) of CO\textsubscript{2} equivalents (CO\textsubscript{2}e)~\cite{luccioni2023estimating,patterson2021carbon,everman2023evaluating}.

\vspace{0.08cm}
We acknowledge that these dimensions and metrics may not be exhaustive, but they offer an overview for evaluating the efficiency and environmental impact of LLM4SE methods. Building on this, we discuss the significance of developing efficient and green LLM4SE techniques and their synergy in the following subsection.

\subsection{Significance of Efficient and Green LLM4SE}

\subsubsection{Why Efficient LLM4SE?}

Most current LLM4SE techniques rely on large-scale models like Code Llama~\cite{roziere2023code}, which incur significant computational costs, requiring vast amounts of FLOPs for both training and inference. According to estimates from the open-source community~\cite{cursorInferenceCharacteristics}, LLMs with architectures similar to Code Llama and LLaMA 2~\cite{touvron2023llama2openfoundation} demand over 100 TeraFLOPs for inference, surpassing the capabilities of many older consumer-grade GPUs~\cite{wikipediaListNvidia}. Running these computationally-intensive LLMs on unsuitable hardware not only results in slow performance, but also risks overheating, potentially damaging the equipment.

Moreover, developing LLM4SE techniques typically involves a prolonged training process of millions of GPU hours~\cite{touvron2023llama,roziere2023code,codegen}. This lengthy training, coupled with the subsequent operation of LLMs, necessitates substantial computational resources, including thousands of high-performance GPUs~\cite{fierceelectronicsUpdateChatGPT} and thousands of GB of memory, translating into high financial costs. For example, training OpenAI's GPT-3~\cite{brown2020languagemodelsfewshotlearners} costs over \$4 million~\cite{cnbcChatGPTGenerative}. Such expenses are prohibitive for many software practitioners, particularly those in smaller companies or developing countries, compelling them to rely on cloud services offered by large or well-funded companies. However, this reliance also brings significant expenses, peaking at \$200,000 per month~\cite{cnbcChatGPTGenerative}, and raises privacy concerns due to the necessity of sharing sensitive data with service providers~\cite{lo2023trustworthy,yang2024robustness}.

Even aside from the costs and privacy issues, users often face a suboptimal experience with cloud-hosted LLM4SE techniques due to prolonged response time caused by network delays~\cite{10.1145/3611643.3616302,compressor,avatar,githubNetwork}. While deploying LLMs on users' devices can address issues of poor user experience from network delays and enhance data privacy~\cite{compressor,avatar,lo2023trustworthy}, accomplishing this task remains challenging without efficient LLM4SE techniques, as most current LLMs demand substantial memory beyond what many personal devices, such as laptops, can accommodate~\cite{10.1145/3611643.3616302,compressor}. Even when memory is sufficient, high inference latency—sometimes lasting several seconds~\cite{10.1145/3611643.3616302,compressor}—and the significant energy consumption of LLMs can drain the battery of personal devices, like laptops, within an hour, making them unsuitable for prolonged use~\cite{10.1145/3611643.3616302,avatar}.
Building efficient LLM4SE techniques can address the above challenges and make LLM4SE techniques more accessible to all software engineering practitioners, regardless of their computational resources.

\subsubsection{Why Green LLM4SE?}
Training LLMs typically requires vast amounts of energy, resulting in high carbon emissions. For example, training LLaMA~\cite{touvron2023llama} consumes 2,638,000 kWh of electricity, equivalent to the annual electricity usage of 445 citizens of Denmark, and emits 1,015 tons of CO\textsubscript{2}e, comparable to the annual emissions of 221 citizens of Denmark~\cite{ieaDenmarkCountries}. The energy consumption of LLM inference is also considerable; each ChatGPT inference consumes 2.9 watt-hours (Wh) of electricity, nearly ten times the 0.3 Wh consumption of a single Google search~\cite{de2023growing,linkedinWhatEnergy}. Moreover, To support these energy-hungry models, data centers and cloud computing facilities—major contributors to global carbon emissions—are heavily relied upon~\cite{belkhir2018assessing,ieeeGenerativeAIs}. These facilities also require enormous amounts of water for cooling and electricity generation, further straining the environment~\cite{li2023making}. In 2022, the combined water usage of data centers powering AI models from companies like Google, Microsoft, and Meta was estimated to be 2.2 billion m\textsuperscript{3}, nearly double Denmark's annual water consumption~\cite{li2023making}. This high energy consumption, water use, and carbon output not only make LLM4SE techniques expensive to develop and maintain but also contribute significantly to climate change and environmental degradation. Therefore, developing more environmentally friendly, green LLM4SE techniques is crucial to mitigate these negative environmental impacts.

% - RAG (Retrieval-Augmented Generation). This pattern became a common response to the costs of retraining LLMs, and it's likely used in LLM4SE applications to reuse the same backend LLM with codebase-dependent code. The authors may want to mention the existence of this pattern and discuss explicitly its impact on the lack of efficiency / green of current LLM4SE techniques.

% - Smaller FLOSS LLMs. The authors highlight how the resource requirements of current LLMs makes most LLM4SE applications rely on cloud-based / API-based LLMs rather than executing it locally. LLaMa (released by Meta), in its 7B parameter configuration, can be run in a single (albeit powerful) consumer-grade machine. The authors could highlight how these smaller FLOSS LLMs are currently providing a potential solution for running LLMs locally.

\subsubsection{Why Synergizing Efficient and Green LLM4SE?}

Efficient and green LLM4SE techniques are closely connected. Shi et al.'s recent study~\cite{avatar} shows that reducing the memory footprint of LLMs can decrease their energy consumption, while both their work and that of Wei et al.~\cite{10.1145/3611643.3616302} also demonstrate that green methods aimed at reducing energy consumption and carbon emissions, such as knowledge distillation and quantization, can also improve the efficiency of LLMs by reducing inference latency and memory usage. These cases illustrate that improving efficiency in LLM4SE techniques can lead to upgraded greenness, and vice versa. However, efficiency and greenness, though related, are not synonymous, meaning that achieving one does not always guarantee the full optimization of the other. For instance, Shi et al.~\cite{compressor} propose an efficient solution by compressing LLMs, but their compressed models still consume significant energy and have unsatisfactory carbon emissions, indicating room for improvement in their environmental impact, which is addressed in their subsequent work~\cite{avatar}. Therefore, we advocate for a synergistic approach that integrates both efficiency and greenness in LLM4SE techniques, achieving the best of both worlds to make LLM4SE techniques both accessible and environmentally sustainable.

% Similarly, Wei et al.~\cite{10.1145/3611643.3616302} apply quantization to LLMs to reduce carbon emissions, while also decreasing inference latency and memory usage. These cases illustrate that improving efficiency in LLM4SE techniques can lead to upgraded greenness, and vice versa. However, efficiency and greenness, though related, are not synonymous, meaning that achieving one does not always guarantee the full optimization of the other. For instance, Shi et al.~\cite{compressor} propose an efficient solution by compressing LLMs, but the compressed models still consume considerable energy and generate significant carbon emissions, indicating room for improvement in their environmental impact, which is addressed in their subsequent work~\cite{avatar}. Therefore, we advocate for a synergistic approach that integrates both efficiency and greenness in LLM4SE techniques, achieving the best of both worlds to make LLM4SE techniques both accessible and environmentally sustainable.
\section{Literature Review}
\label{sec:current_contributions}

\begin{figure}
    \centering
    \includegraphics[width=0.95\textwidth]{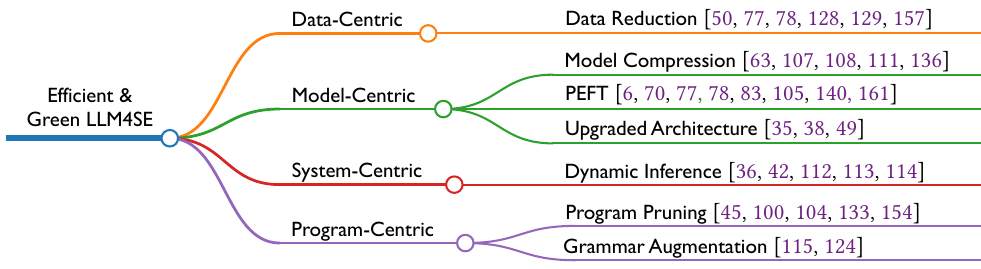}
    \caption{Current Techniques for Efficient and Green LLM4SE}
    \label{fig:llm4se}
\end{figure}

In this section, we review the literature and provide a brief summary of the current state of efficient and green LLM4SE techniques. As shown in~\autoref{fig:llm4se}, we categorize the literature into four main perspectives—data-centric, model-centric, system-centric, and program-centric—based on their methodological focus, and further organize them into specific techniques within each category.

\subsection{Data-Centric Efficient and Green LLM4SE}
The data-centric techniques focus on reducing or optimizing the data required to train LLMs, thereby improving the efficiency and greenness of LLM4SE techniques.

Current data-centric techniques for efficient and green LLM4SE primarily focus on data reduction, which seeks to minimize the amount of data needed to train LLMs, thereby reducing training time, energy consumption, water usage, and carbon emissions. One approach in this direction is active learning, as explored by Hu et al.~\cite{hu2023active}, which apply active learning to CodeBERT and GraphCodeBERT to reduce the size of the fine-tuning data, finding that using less than 10\% of the data significantly degrades LLM performance in code summarization. In contrast, studies~\cite{liu2024delving, wang2023adapter,10298587} have demonstrated that Parameter-Efficient Fine-Tuning (PEFT) methods are more effective with limited data. Notably, Liu et al.~\cite{10298587} show that PEFT can achieve competitive performance in code clone detection with just 1,000 labeled examples. Wang et al.~\cite{wang2022bridging} propose a curriculum learning strategy for training CodeBERT and GraphCodeBERT, where examples are presented in a structured order from simple to complex. This strategy allows their models, which use only 10\% of the training data for code clone detection, to outperform state-of-the-art LLMs trained on the full dataset. More recently, Zhou et al.~\cite{zhou2024programmingexampleliftingpretraining} propose refining and filtering training data for large LLMs using a smaller LLM. By retaining no more than 30\% of the original training data, they achieve up to 20 times lower computational costs in training LLMs like Code Llama, with comparable performance in tasks like code generation and mathematical reasoning.

\subsection{Model-Centric Efficient and Green LLM4SE}

Model-centric techniques focus on optimizing the LLMs themselves, including model parameters and architecture, to improve efficiency and greenness.

\subsubsection{Model Compression.}
Model compression aims to reduce the size of a model, thereby optimizing other relevant metrics such as inference latency, memory usage, and energy consumption. As the pioneering work in compressing code LLMs, Shi et al.~\cite{compressor} propose Compressor, which uses a genetic algorithm-based method for simplifying models and applies knowledge distillation to compress LLMs to 3 megabytes (MB). Their compressed LLMs improve inference latency by 4.23$\times$ on average. Subsequently, Shi et al.~\cite{avatar} also present Avatar, which finds Pareto-optimal compressed LLMs that balance model size, inference latency, energy consumption, and carbon footprint, instead of solely focusing on model size like Compressor. The models produced by Avatar significantly reduce energy consumption (up to 184$\times$ less), carbon footprint (up to 157$\times$ less), and inference latency (up to 76$\times$ faster). Additionally, Su and McMillan~\cite{Su2024} use knowledge distillation on GPT-3.5 to obtain a smaller and more efficient model for code summarization. Wei et al.~\cite{10.1145/3611643.3616302} and Kaushal et al.~\cite{kaushal2023lord} use quantization and low-rank decomposition to compress LLMs for code generation, respectively, both achieving significant efficiency improvements.

\subsubsection{Parameter-Efficient Fine-Tuning (PEFT)}
PEFT aims to improve the training efficiency of LLMs. During training, it updates only a small set of LLM parameters while freezing the rest, thereby greatly reducing computational and memory costs. Researchers have applied PEFT to LLMs for various software engineering tasks~\cite{10298587,liu2024delving,10.1145/3597926.3598036}, such as code generation~\cite{weyssow2024exploring,10.1145/3540250.3558959}, code review~\cite{10299938}, clone detection~\cite{ayupov2022parameterefficient}, and program repair~\cite{10.1145/3691620.3695066}. These studies show that PEFT can significantly reduce the training time and memory consumption of LLMs.

\subsubsection{Upgraded Architecture.}
This line of research focuses on improving the architecture of LLMs to enhance their efficiency. For instance, Hu et al.~\cite{hu2024tackling} propose a novel mechanism in CodeBERT to split long code into short blocks, significantly accelerates the processing of long codes, improving efficiency by over 5$\times$ in code search compared to the original CodeBERT. Gotmare et al.~\cite{10.1145/3611643.3616369} introduce an efficient text-to-code search framework that utilizes cascaded fast and slow LLMs, where a fast model performs rapid retrieval, followed by a slower, larger model for re-ranking to improve accuracy. This framework strikes an optimal balance between efficiency and effectiveness in code search. Additionally, Gu et al.~\cite{gu2022accelerating} present CoSHC, a method designed to accelerate CodeBERT through deep hashing. By leveraging hash codes for code representations generated by CodeBERT, their method reduces retrieval time by over 90\% compared to the original CodeBERT model.

\subsection{System-Centric Efficient and Green LLM4SE}

System-centric techniques focus on optimizing various aspects of the system or pipeline, such as the inference process or decoding strategy, to enhance efficiency and greenness. We list system-centric techniques separately inspired by Miao et al.~\cite{miao2023towards}.

The current techniques in this category mainly focus on applying dynamic inference, which manipulates the inference process of LLMs to improve model efficiency. An early work by Sun et al.~\cite{10172653, 10.1145/3688831} introduces an early rejection mechanism to reject prompts that cannot generate useful completions during inference. This mechanism saves 23.3\% of the computational cost of LLMs in code completion. After that, Sun et al.~\cite{10.1145/3597503.3639120} discover that 54.4\% of tokens can be accurately generated using only the first layer of the GPT-2 model. They develop a method capable of skipping an average of 1.7 layers out of 16 in the models, resulting in a speedup of 11.2\% in code completion. Similarly, Grishina et al.~\cite{10.1145/3611643.3616304} show that only 3 of the 12 layers of CodeBERT are sufficient for vulnerability detection with a 3.3$\times$ speedup in fine-tuning. Recently, Guo et al.~\cite{10.1145/3650212.3680343} introduce CodeFast, a method that terminates the inference process early when excess tokens are unnecessary. This is achieved using a lightweight model called GenGuard, which predicts whether to halt inference at each step. Their method significantly improve the inference speed of various LLMs in code generation, ranging form 34\% to 452\%, without compromising the quality of generated code.

\subsection{Program-Centric Efficient and Green LLM4SE}

Program-centric techniques aim to optimize the input programs fed into LLMs to enhance efficiency and greenness. These methods often focus on reducing the number of tokens or restructuring the grammar of input programs, leading to faster processing and lower resource consumption.

\subsubsection{Program Pruning.}
Pruning the tokens of input programs can improve the efficiency of LLMs, since their inference time is affected by the input length. Zhang et al.~\cite{Zhang2022diet} introduce DietCode, which identifies statements and tokens with the highest attention values, resulting in a 40\% reduction in the computational cost of CodeBERT inference. In addition, Hidvégi et al.~\cite{cigar} optimize prompts to reduce token cost by 62\% in program repair tasks. Shi et al.~\cite{10.1145/3609437.3609438} simplify input for LLMs, resulting in up to 40\% reduction in inference time. Recently, Saad et al.~\cite{saad2024alpine} propose a method to prune unimportant tokens according to the attention scores when token sequences go through the layers of the LLMs, achieving over 58\% reduction memory consumption and 51\% improvements in model throughput. Wang et al.~\cite{10.1145/3643753} introduce SlimCode, which removes redundant tokens by leveraging the structural information of code, such as the Abstract Syntax Tree (AST) and Program Dependency Graph (PDG). This approach improves effectiveness by up to 9\% in code search and summarization, and is up to 133 times faster compared to DietCode.

\subsubsection{Grammar Augmentation.}
Grammar augmentation aims to enhance the efficiency of LLMs by restructuring input programs to reduce token usage, thereby improving inference time. Sun et al.~\cite{10.1145/3650212.3680347} introduce SimPy, the first AI-oriented grammar for Python. SimPy eliminates redundant tokens in Python code, such as unnecessary whitespace and semicolons, using heuristic grammar rules, while preserving the original AST structure. This approach reduces token usage by 13.5\% and improves performance by 10.4\% in code-related tasks for models like Code Llama and GPT-4. Additionally, Ugare et al.~\cite{ugare2024improving} propose SynCode, which integrates a context-free grammar into the decoding process of LLMs. This method forces the model to efficiently generate syntactically valid code, accelerating the decoding process by up to 19\% when generating syntactically correct code snippets.

% ~\cite{barnett2024seven}

\vspace{0.2cm}

In addition to the above techniques, several studies do not provide a specific technique but focus on the efficiency and greenness of the code generated by LLM4SE techniques, such as~\cite{liu2023refining, huang2024effibench, vartziotis2024learn}. While these studies make valuable contributions to the LLM4SE area, they are not included in the above categories due to their primary focus on the generated code rather than the LLMs themselves, which is beyond the scope of our literature review.

Additionally, several studies focus on the greenness of machine learning/deep learning models outside the LLM4SE domain, such as~\cite{duran2024identifying,10.1145/3644815.3644942,alizadeh2024green,husom2024engineering,yuan2024impact}. Among these, Duran et al.~\cite{duran2024identifying} analyze ML serving architectural design decisions and their impact on energy efficiency. Kannan et al.~\cite{10.1145/3644815.3644942} introduce GreenRunner, a tool that selects models based on a natural language description of the task, with an emphasis on energy efficiency. Alizadeh et al.~\cite{alizadeh2024green} examine the energy efficiency and inference performance of DL runtime infrastructures across frameworks such as PyTorch and TensorFlow. Husom et al.~\cite{husom2024engineering} propose CEMAI, a tool designed to monitor and analyze carbon emissions across the entire lifecycle of ML model development. Yuan et al.~\cite{yuan2024impact} empirically assess whether knowledge distillation can improve the energy efficiency of LLMs in natural language processing domains without compromising performance. These studies are excluded from our review, as they do not specifically address the efficiency and greenness of LLM4SE techniques. However, their findings could be valuable for future research in this area. For example, adapting Kannan et al.'s ~\cite{10.1145/3644815.3644942} GreenRunner tool to select energy-efficient LLMs for software engineering tasks may be beneficial, and leveraging Duran et al.'s~\cite{duran2024identifying} architectural design decisions potentially helps optimize the energy efficiency of LLM4SE techniques. We encourage future research to explore these possibilities.

\find{{\bf Reflection ---} The literature review highlights remarkable progress in developing efficient and green LLM4SE techniques across data, model, system, and program-centric perspectives. However, this area remains far from fully explored, with more breakthroughs needed to address the efficiency and greenness challenges of LLM4SE.}
\section{Future: LLM4SE of 2030 and Beyond}
\label{sec:vision}

After reviewing the current state of efficient and green LLM4SE solutions, we now turn our attention to the future. According to the current trends and the potential of our research community, we envision a future where efficient and green LLM4SE solutions revolutionize the LLM-based software engineering tool landscape, benefiting various stakeholders, including industry, individual practitioners, and society.

\subsection{For Individual Software Practitioners}
The advancement of efficient and green LLM4SE can pave the way for the emergence and widespread adoption of {\it Private, Personalized, Trusted, and Collaborative Software Engineering Assistants}. These new assistants will transcend current tools such as GitHub Copilot~\cite{githubGitHubCopilot}, which are confined to one single task like code suggestion, rely on cloud infrastructure, pose potential privacy risks~\cite{yang2023memorzation,niu2023codexleaks}, and are challenging or costly to customize~\cite{githubCustomizingFinetuning}. With minimal computational resources and energy consumption, these new assistants can be seamlessly integrated into software practitioners' local development environments. They remain entirely private to the practitioners, securing their intellectual property and sensitive information from various threats~\cite{debenedetti2023privacy,yang2023memorzation,niu2023codexleaks,huang2023not}. This integration facilitates real-time interaction between software practitioners and their assistants with minimal latency, enabling them to provide instant feedback and suggestions to practitioners as they write code, test software, or engage in other software engineering activities. Such real-time interactions can occur without an Internet connection, which is particularly crucial in regions of developing countries with limited access to high-speed Internet. Furthermore, this interaction can consume less computational resources and energy than today, making it financially viable for individual practitioners to utilize these assistants on personal devices like laptops.

Moreover, the assistants would deliver personalized assistance for various software engineering tasks, tailored to individual preferences and project requirements. Such personalization benefits from the efficient and green nature of the assistants, allowing it to be accomplished with minimal computational resources and energy consumption. Following such personalization, the assistants will furnish more precise and pertinent assistance to software practitioners, thereby catalyzing trust between software practitioners and their assistants. This trust will enhance the collaboration between software practitioners and assistants, enabling them to work together more effectively to improve productivity, software quality, and the overall development experience.

\subsection{For Industry}
Advances in efficient and green LLM4SE will facilitate the development of {\it Low-cost and Low-latency LLM-based Software Engineering Tools}. The existing LLM-based software engineering tool landscape is dominated by heavyweight solutions that are computationally expensive and energy-intensive. Companies must not only invest in large amounts of computing resources such as GPUs to develop and run these tools, but also have to bear the high cost of the associated energy consumption (electricity, cooling, etc.) and carbon emissions, requiring the purchase of many carbon credits to compensate for the environmental impact~\cite{carboncreditsMetaAlmost}. These costs limit the existing tools' profitability. For example, Microsoft's GitHub Copilot currently loses \$20 per user per month instead of making a profit~\cite{aibusinessMicrosoftsGitHub}. Moreover, these tools often have high latency, often resulting in a poor user experience~\cite{githubNetwork,10.1145/3611643.3616302}.

By adopting efficient and green LLM4SE solutions, companies can replace existing imperfect tools with upgraded ones that are faster, greener, and require fewer resources to develop and operate. Not only can these upgraded tools be used internally by companies to streamline their software development processes, but they can also be offered as Software as a Service (SaaS) solutions to other companies and developers, giving them access to cutting-edge LLM-based software engineering tools at reduced costs and faster response times, all without requiring a large investment in computing resources. In addition, efficient and green LLM4SE solutions can diminish the entry barriers for LLM-based software engineering tools, presenting significant opportunities for Small and Medium Enterprises (SMEs) and startups, which typically face resource constraints while striving to improve their development workflows. This fosters a more equitable environment in the software industry, allowing SMEs and startups to compete on par with larger companies. Overall, for the software industry, adopting efficient and green LLM4SE will revolutionize the LLM-based software engineering tool landscape, making it more affordable and accessible to all stakeholders within the industry.

\subsection{For Society}
Efficient and green LLM4SE will play a critical role in fostering {\it Better Environmental Sustainability in Software Engineering}. The software industry is a significant contributor to global carbon emissions and energy consumption, with data centers and cloud computing facilities, often used for LLM development and operations, being major responsible parties~\cite{belkhir2018assessing,ieeeGenerativeAIs}. Embracing efficient and green LLM4SE solutions can mitigate these environmental impacts. These new solutions promise to reduce the energy consumption and carbon emissions associated with LLM development and operation, thereby reducing reliance on data centers and cloud computing facilities and promoting greener practices in the software industry.
Transitioning to efficient and green LLM4SE will not only benefit the environment, but will also align with the growing societal emphasis on environmental sustainability. Large technology companies can achieve sustainability goals such as carbon neutrality~\cite{wang2021technologies,sustainabilityAimingAchieve}, while smaller companies and individual developers can share more carbon credits, facilitating their rapid growth~\cite{carboncreditsMetaAlmost}. By promoting efficient and green LLM4SE, the software industry can demonstrate its commitment to environmental responsibility and contribute to a more sustainable future for all.

\find{{\bf Vision ---} By 2030, the future landscape of LLM4SE tools will feature low-cost assistants integrated into local development environments, offering real-time support to practitioners, while industry-wide adoption of greener tools will reduce environmental impact, promote sustainability, and ensure equitable access for all stakeholders, benefiting both the software industry and society.}
\section{The Road Ahead}
\label{sec:road}

\begin{figure}
    \centering
    \includegraphics[width=0.8\columnwidth]{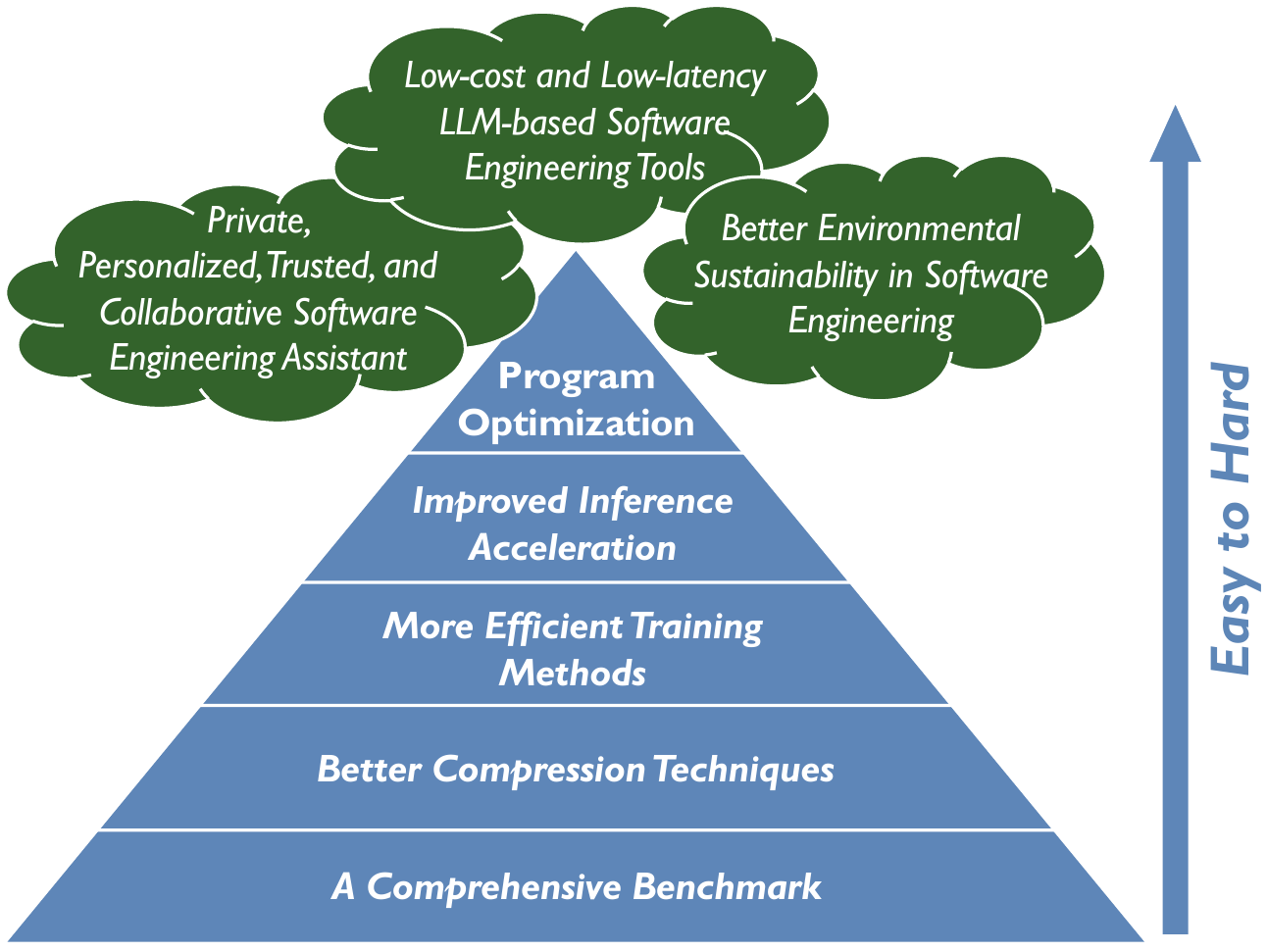}
    \caption{Roadmap for Efficient and Green LLM4SE}
    \label{fig:roadmap}
\end{figure}

This section presents a roadmap for achieving efficient and sustainable LLM4SE, outlining several specific research paths and potential techniques for further exploration by the research community. It is important to note that some research paths may overlap or be interrelated, and may not fit exclusively within the categories of data, model, system, or program-centric approaches discussed in~\autoref{sec:current_contributions}. Therefore, the roadmap is presented holistically, with each research topic or technique potentially addressing multiple dimensions. \autoref{fig:roadmap} illustrates this roadmap for efficient and green LLM4SE. Based on the research gaps identified in~\autoref{sec:current_contributions}, we sort the roadmap's difficulty level from low to high, according to the current maturity of that research areas. If a research path has fewer studies or is more challenging to address, we classify it as high difficulty.

\subsection{A Comprehensive Benchmark}
Prior to embarking on the development of LLM4SE techniques, it is essential to establish a comprehensive benchmark dataset to evaluate their efficiency and greenness. While many existing benchmarks, such as CodeXGLUE~\cite{CodeXGLUE} and CoderEval~\cite{yu2024codereval}, evaluate the effectiveness (e.g., accuracy) of LLMs in certain software engineering tasks, very few specifically target the efficiency and greenness of LLM4SE. A few studies have collected datasets to evaluate the efficiency or energy consumption of the code generated by LLMs, such as EffiBench~\cite{huang2024effibench}, and the datasets constructed by Liu et al.~\cite{liu2023refining} and Vartziotis~\cite{vartziotis2024learn}. Nevertheless, these datasets primarily focus on evaluating the generated code rather than the LLMs, and thus are not comprehensive enough. Consequently, researchers and practitioners lack a comprehensive understanding of the current landscape of LLM4SE development in terms of efficiency and greenness, making it difficult for them to compare different LLM4SE techniques and make informed decisions about which one to adopt. We suggest that the research community prioritize curating a diverse set of tasks and datasets representative of real-world software engineering applications to benchmark the efficiency and greenness of LLM4SE techniques. These efforts will lay the foundation for future advances in the area.

\subsection{More Efficient Training Methods}
Training LLMs is computationally intensive and time-consuming, posing a significant bottleneck in the development of LLM4SE. While some studies have leveraged PEFT-based methods to accelerate LLM training, as discussed in~\autoref{sec:current_contributions}, more efficient training methods, such as data and model parallelism~\cite{zhao2024galore}, pipeline parallelism~\cite{kim2023bpipe}, and improved optimizers~\cite{you2019large}, have not been thoroughly explored for LLM4SE. Data parallelism involves distributing data across multiple devices, with each device holding a full copy of the model, while model parallelism splits the model across devices, with each device holding only a portion. Both methods can accelerate LLM training when multiple GPUs are available. Pipeline parallelism combines data and model parallelism, making it suitable for training very large LLMs. Additionally, You et al.~\cite{you2019large} proposed a novel optimizer that automatically adjusts the learning rate for LLMs' each layer and batch size, enabling the use of very large batch sizes for faster convergence of LLMs. This optimizer allowed for pre-training a BERT model in just 76 minutes back in 2019, without relying on large GPUs. However, no study has yet systematically evaluated these methods in the context of LLM4SE. We suggest that researchers apply and assess these approaches to identify the most effective techniques and the unique challenges in developing LLM4SE, ultimately paving the way for new training acceleration methods tailored specifically for LLM4SE.

Furthermore, Retrieval-Augmented Generation (RAG)~\cite{lewis2020retrieval} could be explored as a way to reduce the cost of specializing LLMs for software engineering tasks. RAG retrieves texts that semantically match a query from an external knowledge base and passes them to an LLM to extract the correct answer, allowing LLMs to leverage existing knowledge to generate domain-specific responses without requiring extensive retraining~\cite{fan2024survey}. While recent studies have demonstrated RAG's effectiveness in software engineering tasks such as code suggestion~\cite{codesuggestion,wang2024coderag} and summarization~\cite{parvez2021retrieval}, its primary benefit for LLM4SE lies in eliminating the need for fine-tuning or training LLMs from scratch. However, the inference cost of RAG remains high, even surpassing that of standalone LLMs due to the additional queries made to the knowledge base~\cite{hofstatter2023fid}. Therefore, we recommend that researchers explore ways to optimize RAG's inference cost for LLM4SE, such as using more efficient retrieval methods~\cite{izacard2020leveraging, hofstatter2023fid} or caching retrieved knowledge to reduce the number of queries to the knowledge base~\cite{jin2024ragcache}.

\subsection{Better Compression Techniques}
Popular model compression techniques, including knowledge distillation and quantization, have shown promise in compressing LLMs for efficient and green LLM4SE, as introduced in~\autoref{sec:current_contributions}. However, further exploration is still needed to optimize the efficiency of LLMs to a satisfactory extent, aiming at response times in the range of a few tens of milliseconds~\cite{aye2020sequence} and memory consumption of less than 50 MB~\cite{svyatkovskiy2021fast}. Current compression techniques may fall short of these requirements, especially for very large LLMs like Code Llama~\cite{roziere2023code}, which can range from tens to hundreds of gigabytes in size. Therefore, there is a pressing need to propose novel compression techniques tailored to these very large LLMs in order to achieve the desired efficiency. Furthermore, while knowledge distillation has been the primary focus of current compression efforts for LLM4SE techniques~\cite{compressor,avatar,Su2024}, other techniques like quantization and pruning have received relatively less attention, with fewer studies available. Hence, we advocate for increased research efforts to explore the potential of quantization and pruning techniques alongside the innovation of new knowledge distillation methods. Moreover, combining these techniques can also be investigated to achieve better efficiency and greenness improvements for LLM4SE techniques.

\subsection{Improved Inference Acceleration}

Similar to training, the inference process of LLMs is computationally demanding, creating another bottleneck in adopting LLM4SE. While some studies have proposed dynamic inference methods to accelerate LLM inference~\cite{10.1145/3597503.3639120, 10172653}, comprehensive exploration in this area remains limited. For example, cascade inference strategies, which arrange small and large LLMs in a cascading manner and adaptively select suitable models rather than relying solely on large, slow LLMs for every query, could be investigated for LLM4SE. Additionally, methods such as non-autoregressive decoding~\cite{ghazvininejad2019mask,gu2018non} and speculative decoding~\cite{chen2023accelerating,leviathan2023fast} may also accelerate LLM inference and have potential applicability to LLM4SE~\cite{miao2023towards}. Among these methods, non-autoregressive decoding speeds up LLM inference by generating all tokens in the sequence simultaneously, instead of sequentially as in auto-regressive decoding. Speculative decoding also improves the efficiency of model inference by using a smaller, faster draft model to generate text first, with the large LLM only verifying and adjusting tokens if necessary. If any token's logit from the large LLM differs substantially from the draft model's prediction, the large LLM regenerates that token. This process requires just one forward operation from the large LLM, significantly reducing computation time. We recommend that researchers systematically evaluate the effectiveness of these approaches in the context of LLM4SE and develop more efficient inference acceleration techniques tailored to these applications.

Furthermore, a recent study by Wei et al.~\cite{wei2023copiloting} leverages code completion engines to filter out infeasible tokens generated by LLMs. By integrating lightweight static code completion engines to directly complete tokens without querying the LLM when only one valid option exists, efficiency in program repair tasks is improved. This highlights the potential of combining LLMs with existing program analysis tools to enhance both efficiency and sustainability in LLM4SE, offering another promising direction for further research.

\subsection{Program Optimization}
One area not touched in LLM4SE is optimizing the programs built for LLM inference. The inference pipeline of LLMs is typically constructed using human-written code, which may not fully exploit specific hardware features such as CPU/GPU instructions for parallel computing~\cite{li2023sirius}. To address this, researchers can develop program transformation tools that automatically transform LLM inference code to take advantage of specific backend libraries or hardware features~\cite{miao2023towards}. In addition, direct optimization of the LLM inference code by eliminating dead code~\cite{theodoridis2022finding} and tuning compiler optimization flags~\cite{chen2021efficient} can also improve the efficiency and energy consumption of LLM4SE techniques. We recommend that the software engineering community further explore these techniques to help LLM inference achieve better efficiency and lower environmental impact in LLM4SE techniques.

Moreover, as a complementary direction to the existing work on specialized program grammars for LLMs, such as SimPy~\cite{10.1145/3650212.3680347}, we suggest that researchers also focus on developing optimization techniques for these emerging grammars to further enhance the efficiency and sustainability of LLM4SE techniques. For example, efforts to verify the correctness of transformation rules~\cite{cheng2020certifying} and to merge or eliminate redundant rules~\cite{steinhofel2020refinity} could reduce the complexity of program grammars. These optimizations would not only streamline the grammar but also improve LLMs' efficiency in code generation, contributing to more environmentally friendly LLM4SE techniques.

\find{{\bf Roadmap ---} Our research community can advance towards more efficient and green LLM4SE by developing comprehensive benchmarks, exploring more efficient training methods, optimizing compression and inference acceleration techniques, and enhancing program-level optimizations to reduce computational and environmental costs in LLM-based software engineering tools.}

\section{Conclusion and Future Work}
\label{sec:conclusion}

As LLMs gain traction in the software engineering community, concerns about their efficiency (i.e., computational cost, memory usage, and time consumption) and greenness (i.e., energy usage, water consumption, and carbon emissions) have escalated. This paper provides a brief overview of the current landscape of LLM4SE solutions, focusing on their efficiency and greenness while highlighting the associated challenges and opportunities in this emerging field. We then sketch a future vision to provide insights into the potential benefits of adopting efficient and green LLM4SE solutions for industry, individual practitioners, and society. To realize this vision, we propose a roadmap for future research, outlining specific directions and potential solutions that the research community can pursue. With these efforts, we aim to motivate more people to join and contribute to the LLM4SE research journey, with the goal of fostering a new era of software engineering where LLM4SE solutions are not only effective, but also efficient and environmentally sustainable.

\begin{acks}
  This research / project is supported by the National Research Foundation, under its Investigatorship Grant (NRF-NRFI08-2022-0002). Any opinions, findings and conclusions or recommendations expressed in this material are those of the author(s) and do not reflect the views of National Research Foundation, Singapore.
\end{acks}

\bibliographystyle{ACM-Reference-Format}
\bibliography{references}

\end{document}